\documentclass[hyphens, nofootinbib, twocolumn]{revtex4}

\usepackage{amsmath}
\usepackage{amssymb}
\usepackage{color}
\usepackage{graphicx}
\usepackage[utf8]{inputenc}
\usepackage{url}

\usepackage[bookmarks, bookmarksopen, bookmarksnumbered]{hyperref}
\usepackage[all]{hypcap}

\begin{document}

\title{Time Step Size Limitation Introduced by the BSSN Gamma Driver}

\author{Erik Schnetter}
\affiliation{Center for Computation \& Technology, Louisiana State
  University, USA}
\affiliation{Department of Physics \& Astronomy, Louisiana State
  University, USA}
\email{schnetter@cct.lsu.edu}
\homepage{http://www.cct.lsu.edu/~eschnett/}

\date{March 3, 2010}

\begin{abstract}
  Many mesh refinement simulations currently performed in numerical
  relativity counteract instabilities near the outer boundary of the
  simulation domain either by changes to the mesh refinement scheme or
  by changes to the gauge condition.  We point out that the BSSN
  \emph{Gamma Driver} gauge condition introduces a time step size
  limitation in a similar manner as a CFL condition, but which is
  independent of the spatial resolution.  We give a didactic
  explanation of this issue, show why especially mesh refinement
  simulations suffer from it, and point to a simple remedy.
\end{abstract}

\maketitle

\section{Introduction}

We assume that the audience is familiar with the concept of a
Courant-Friedrichs-Lewy (CFL) condition
\cite{wikipedia-CFL-condition}.  Loosely speaking, the CFL condition
states: When a partial differential equation, for example the wave
equation
\begin{eqnarray}
  \label{eq:wave}
  \partial_t^2 u & = & c^2\, \Delta u \quad\textrm{,}
\end{eqnarray}
is integrated numerically, then the time step size $\delta t$ is
limited by the spatial resolution $\delta x$ and the maximum
propagation speed $c$ by
\begin{eqnarray}
  \label{eq:cfl}
  \delta t & < & Q\, \frac{\delta x}{c} \quad\textrm{.}
\end{eqnarray}
Here $Q$ is a constant of order $1$ that depends on the time
integration method (and details of the spatial discretisation).
Choosing a time step size larger than this is unstable and must
therefore be avoided.  (There are time integration methods that do not
have such a stability limit, but these are expensive and not commonly
used in numerical relativity, so we will ignore them here.)

%

\section{Example: Exponential Decay}

In real-world equations, there are also other restrictions which limit
the time step size, and which may be independent of the spatial
resolution.  One simple example for this is the exponential decay
\begin{eqnarray}
  \label{eq:decay}
  \partial_t u & = & - \lambda\, u
\end{eqnarray}
where $\lambda > 0$ is the decay constant.  Note that this equation is
an ordinary differential equation, as there are no spatial
derivatives.  The solutions of (\ref{eq:decay}) are given by
\begin{eqnarray}
  u(t) & = & A\, \exp\{ - \lambda t \}
\end{eqnarray}
with amplitude $A$.

The decay constant $\lambda$ has dimension $1/T$.  The time step size
is limited by
\begin{eqnarray}
  \delta t & < & Q'\, \frac{1}{\lambda}
\end{eqnarray}
where $Q'$ is a constant of order $1$ that depends on the time
integration method.  Choosing a time step size larger than this is
unstable and must therefore be avoided.  (As with the CFL criterion,
there are time integration methods that do not have such a stability
limit.)

As an example, let us consider the forward Euler scheme with a step
size $\delta t$.  This leads to the discrete time evolution equation
\begin{eqnarray}
  \frac{u^{n+1} - u^n}{\delta t} & = & - \lambda\, u^n
\end{eqnarray}
or
\begin{eqnarray}
  u^{n+1} & = & (1 - \delta t\, \lambda)\, u^n \quad\textrm{.}
\end{eqnarray}
This system is unstable e.g.\ if $|u^{n+1}| > |u^n|$ (there are also
other definitions of stability), or if
\begin{eqnarray}
  |1 - \delta t\, \lambda| & > & 1 \quad\textrm{,}
\end{eqnarray}
which is the case for $\delta t > 2 / \lambda$ (and also for $\delta t
< 0$).  In this case, the solution oscillates between positive and
negative values with an exponentially growing amplitude.

\section{Gamma Driver}

The BSSN \cite{Alcubierre99d} Gamma Driver condition is a time
evolution equation for the shift vector $\beta^i$, given by (see e.g.\
(43) in \cite{Alcubierre02a})
\begin{eqnarray}
  \label{eq:gamma-driver}
  \partial_t^2 \beta^i & = & F\, \partial_t \tilde \Gamma^i -
  \eta\, \partial_t \beta^i \quad\textrm{.}
\end{eqnarray}
There exist variations of the Gamma Driver condition, but the
fundamental form of the equation remains the same.  The term
$F\, \partial_t \tilde \Gamma^i$ contains second spatial derivatives
of the shift $\beta^i$ and renders this a hyperbolic, wave-type
equation for the shift.  The parameter $\eta>0$ is a damping
parameter, very similar to $\lambda$ in (\ref{eq:decay}) above.  It
drives $\partial_t \beta^i$ to zero, so that the shift $\beta^i$ will
tend to a constant in stationary spacetimes.  (This makes this a
\emph{symmetry-seeking} gauge condition, since $\partial_t$ will then
tend to the corresponding Killing vector.)

Let us now consider a simple spacetime which is spatially homogeneous,
i.e.\ where all spatial derivatives vanish.  In this case (see e.g.\
(40) in \cite{Alcubierre02a}), $\partial_t \tilde \Gamma^i = 0$, and
only the damped oscillator equation
\begin{eqnarray}
  \partial_t^2 \beta^i & = & - \eta\, \partial_t \beta^i
\end{eqnarray}
remains.  As we have seen above, solving this equation numerically
still imposes a time step size limit, even though there is no length
scale introduced by the spatial discretisation, so the spatial
resolution can be chosen to be arbitrarily large; there is therefore
no CFL limit.  This demonstrates that the damping time scale set by
the parameter $\eta$ introduces a resolution-independent time step
size limit.

This instability was e.g.\ reported in \cite{Sperhake:2006cy}, below
(13) there, without explaining its cause.  The authors state that the
choice $\eta=2$ is unstable near the outer boundary, and they
therefore choose $\eta=1$ instead.  Decreasing $\eta$ by a factor of
$2$ increases the time step size limit correspondingly.

The explanation presented above was first brought forth by Carsten
Gundlach \cite{Gundlach2008a} and Ian Hawke \cite{Hawke2008a}.  To our
knowledge, it has not yet been discussed in the literature elsewhere.

Harmonic formulations of the Einstein equations have driver parameters
similar to the BSSN Gamma Driver parameter $\eta$.  Spatially varying
parameters were introduced in harmonic formulations to simplify the
gauge dynamics in the wave extraction zone far away from the origin
(see e.g.\ (8) in \cite{Scheel:2008rj}).  \cite{Palenzuela:2009hx}
uses a harmonic formulation with mesh refinement, and describes using
this spatial dependence also to avoid time stepping instabilities (see
(45) there).

\section{Mesh Refinement}

When using mesh refinement to study compact objects, such as black
holes, neutron stars, or binary systems of these, one generally uses a
grid structure that has a fine resolution near the centre and
successively coarser resolutions further away from the centre.  With
full Berger-Oliger AMR that uses sub-cycling in time, the CFL factors
on all refinement levels are the same, and thus the time step sizes
increase as one moves away from the centre.  This makes it possible
that the time step size on the coarsest grids does not satisfy the
stability condition for the Gamma Driver damping parameter $\eta$ any
more.

One solution to this problem is to omit sub-cycling in time for the
coarsest grids by choosing the same time step size for some of the
coarsest grids.  This was first advocated by \cite{Bruegmann:2003aw},
although it was introduced there to allow large shift vectors near the
outer boundary as necessary for a co-rotating coordinate system.  It
was later used in \cite{Brugmann:2008zz} (see section IV there) to
avoid an instability near the outer boundary, although the instability
is there not attributed to the Gamma Driver.  Omitting sub-cycling in
time on the coarsest grids often increases the computational cost only
marginally, since most of the computation time is spent on the finest
levels.

Another solution is to choose a spatially varying parameter $\eta$,
e.g.\ based on the coordinate radius and mimicking the temporal
resolution of the grid structure, which may grow linearly with the
radius.  This follows the interpretation of $\eta$ setting the damping
timescale, which must not be larger than the timescale set by the time
discretisation.

One possible spatially varying definition for $\eta$ could be
\begin{eqnarray}
  \label{eq:varying}
  \eta(r) & := & \eta^*\; \frac{R^2}{r^2 + R^2} \quad,
\end{eqnarray}
where $r$ is the coordinate distance from the centre of the black
hole.  The parameter $R$ defines a transition radius between an inner
region, where $\eta$ is approximately equal to $\eta^*$, and an outer
region, where $\eta$ gradually decreases to zero.  This definition is
simple, smooth, and differentiable, and mimics a ``typical'' mesh
refinement setup, where the resolution $h$ grows approximately
linearly with the radius $r$.

Another, simpler definition for $\eta$ (which is not smooth -- but
smoothness is not necessary; $\eta$ could even be discontinuous) is
\begin{eqnarray}
  \label{eq:varying-simple}
  \eta(r) & := & \eta^*\; \left\{
    \begin{array}{llll}
      1 & \mathrm{for} & r \le R & \textrm{(near the origin)}
      \\
      \frac{R}{r} & \mathrm{for} & r \ge R & \textrm{(far away)}
    \end{array}
    \right. ,
\end{eqnarray}
which is e.g.\ implemented in the \texttt{McLachlan} code
\cite{ES-mclachlanweb}.

If there are multiple black holes, possibly with differing resolution
requirements, then prescriptions such as (\ref{eq:varying}) or
(\ref{eq:varying-simple}) need to be suitably generalised, e.g.\ via
\begin{eqnarray}
  \label{eq:multiple}
  \frac{1}{\eta(r)} & := & \frac{1}{\eta_1(r_1)} +
  \frac{1}{\eta_2(r_2)} \quad,
\end{eqnarray}
where $\eta_1$ and $\eta_2$ are the contributions from the individual
black holes, with $r_1$ and $r_2$ the distances to their centres.
This form of (\ref{eq:multiple}) is motivated by the dimension of
$\eta$, which is $1/M$, so that two superposed black holes of masses
$m_1$ and $m_2$ lead to the same definition of $\eta$ as a single
black hole with mass $m_1+m_2$.

Another prescription for a spatially varying $\eta$ has been suggested
in \cite{Mueller:2009jx}.  In this prescription, $\eta$ depends on the
determinant of the three-metric, and it thus takes the masses of the
black hole(s) automatically into account.  This prescription is
motivated by binary systems of black holes with unequal masses, where
$\eta$ near the individual black holes should be adapted to the
individual black holes' masses, and it may be more suitable to use
this instead of (\ref{eq:multiple}).

There can be other limitations of the time step size near the outer
boundary, coming e.g.\ from the boundary condition itself.  In
particular, radiative boundary conditions impose a CFL limit that may
be stricter than the CFL condition from the time evolution equations
in the interior.

\begin{acknowledgements}
  We thank Peter Diener, Christian D. Ott, and Ulrich Sperhake for
  valuable input, and for suggesting and implementing the spatially
  varying beta driver in (\ref{eq:varying-simple}).  We also thank
  Bernd Brügmann for his comments.
  This work was supported by the NSF awards \#0721915 and \#0905046.
  It used computational resources provided by LSU, LONI, and the NSF
  TeraGrid allocation TG-MCA02N014.
\end{acknowledgements}

\bibliographystyle{bibtex/unsrt-url}
\bibliography{bibtex/references,publications-schnetter,local}

\end{document}